\documentstyle[aps,prd]{revtex}
\renewcommand{\baselinestretch}{1.2}
\begin{document}
\large
\title{Leptons in Dirac Spin Networks}
\author{Walter Smilga}
\address{Isardamm 135 d, D-82538 Geretsried, Germany}
\address{e-mail: wsmilga@compuserve.com}
\maketitle
\renewcommand{\baselinestretch}{1.2}

\begin{abstract}
In large networks of Dirac spinors individual spinors show space-time
properties relative to quasi-classical clusters of spinors. 
Three forms of relations between spinors and such clusters
are identified. 
These constitute three families of particle-like configurations, with a 
mass spectrum in close agreement with the experimental lepton spectrum.
\end{abstract}


\renewcommand{\baselinestretch}{1.0}

\section{Introduction}

In 1971 R.~Penrose \cite{rp} made an attempt to describe the geometry of 
space-time in a purely combinatorial way. 
Penrose studied networks of two-component spinors, which represented the 
simplest quantum mechanical objects. 
He was able to show that large clusters of such spinors, generated properties 
of angular direction in three-dimensional space. 
Despite this success, the concept of SU(2) based spin networks was not 
considered rich enough to also describe distance \cite{rpwr}. 
Penrose's interest, therefore, turned to more complex twistor objects.
Nowadays various kinds of spin networks are discussed as models of space-time 
at Planck scales \cite{jcb}.

This article is based on Dirac spinors. 
Dirac spinors are understood as four-component spinors. They transform under 
the operations of the homogeneous Lorentz group by the application of 
$4\times4$-matrices, which are generated from well-known Dirac matrices. 
They do not have space-time properties, nor do they carry energy-momentum in 
the usual sense.
Two Dirac spinors can be distinguished only by their spin. 
Two spinors in the same spin state are indistinguishable.

A network of Dirac spinors is a set of Dirac spinors, and an instruction of 
how to link together spinors, to form clusters with well-defined 
quantum numbers.
The linking of spinors can rapidly lead to complex structures.
This article, therefore, analyses some structures of physical interest 
within very large networks of Dirac spinors.
In doing so it takes advantage of laws of large numbers and of symmetry 
rules.
 
An elementary rule of quantum mechanics will be used to identify substructures
within Dirac spin networks:
If a state is a product of two states, corresponding to two subsets of 
spinors, then the two structures are separable, and can, therefore, be treated 
as independent quantum mechanical systems. 
Otherwise the states are \begin{em}entangled\end{em}, and the structure has 
to be treated as a coherent quantum mechanical system.

\section{ Dirac spinors and de~Sitter group}

Dirac spinors can be represented by linear combinations with complex
coefficients of four-component basic vectors:

\begin{eqnarray}
|u_a\rangle &:=& \left(\begin{array}{c} 1\\0\\0\\0 \end{array}\right)
 \; ,\;\;\;\;\;
|u_b\rangle  :=  \left(\begin{array}{c} 0\\1\\0\\0 \end{array}\right)
\; , \;                                             \label{3-2a}\\
\nonumber \\
|v_a\rangle &:=& \left(\begin{array}{c} 0\\0\\0\\1 \end{array}\right)
 \; , \;\;\;\;\;
|v_b\rangle  :=  \left(\begin{array}{c} 0\\0\\1\\0 \end{array}\right) 
\; .                                                \label{3-2b}
\end{eqnarray}

Dirac spinors form a complex linear vector space, which can easily be 
extended to a Hilbert space $H$ by adding the scalar product 
\begin{equation}
\langle \bar{a} | b \rangle \; \mbox{ with } \;
\langle \bar{a} | = \langle a | \gamma^0  \; ,                  \label{3-8}
\end{equation}
which is well-known from Dirac's theory of electrons.

Dirac's $\gamma$-matrices generate symmetry transformations in $H$.
They satisfy the anti-commutation relations
\begin{equation}
\{\gamma_\mu, \gamma_\nu \} = 2 g_{\mu\nu}                      \label{3-7a}
\end{equation}
and the commutation relations
\begin{equation}
\frac{i}{2} \, [\gamma_\mu, \gamma_\nu ] = \sigma_{\mu\nu}\;,   \label{3-7b}
\end{equation}
where $\mu, \nu = 0,\ldots,3$.

The $4\times4$-matrices
\begin{equation}
s_{\mu\nu} :=\, \frac{1}{2} \sigma_{\mu\nu}
\; \mbox{ and } \;
s_{\mu4} :=\, \frac{1}{2} \gamma_\mu \; ,                       \label{3-9}
\end{equation}
form an irreducible representation of the \begin{em}de~Sitter group\end{em} 
SO(3,2) on the Hilbert space of Dirac spinors.
The proof of this is by inserting (\ref{3-9}) into the commutation relation 
of SO(3,2)
\begin{equation}
[s_{\mu\nu}, s_{\rho\sigma}] =
-i[g_{\mu\rho} s_{\nu\sigma} - g_{\mu\sigma}
s_{\nu\rho} + g_{\nu\sigma} s_{\mu\rho}
- g_{\nu\rho} s_{\mu\sigma}] \; ,                               \label{3-10}
\end{equation}
\begin{equation}
[s_{\mu4}, s_{\nu4}] = -i s_{\mu\nu} \; ,                       \label{3-11}
\end{equation}
\begin{equation}
[s_{\mu\nu}, s_{\rho4}]
= i[g_{\nu\rho} s_{\mu4} - g_{\mu\rho} s_{\nu4}] \; .           \label{3-12}
\end{equation}

The Hilbert space $H$ of Dirac spinors, equipped with symmetry transformations
generated by the $4\times4$-matrices (\ref{3-9}), will be the basis for 
this article.

A system of $N$ Dirac spinors is then described by states in a product
Hilbert space $H*...*H$.
By reducing product representations with respect to SO(3,2), new 
representations are obtained with different spin quantum numbers.
This process links spinor states to form product states with defined quantum 
numbers. This process is well-known from quantum mechanics of angular
momentum. 
Within this article a set of $N$ Dirac spinors, described by means of a 
Hilbert space $H*...*H$, and equipped with this linkage will be called 
a \begin{em}Dirac spin network\end{em}.

\section{Poincar\'{e} group and space-time}

Consider a large number of, say $N = 10^{23}$, Dirac spinors, 
forming product states with large quasi-continuous total quantum numbers.
Such a cluster will be called a \begin{em}macro-object\end{em} or simply
\begin{em}object\end{em} and its state a \begin{em}macro-state\end{em} 
for short.

The generators $S_{ab}$ of SO(3,2)-transformations of the macro-objects are 
defined by the sum over the corresponding operators $s_{ab}$ of the individual
spinors.

In the following, the transformations generated by $S_{\mu4}$ will be 
restricted to infinitesimal small transformations.
This can be expressed by replacing $S_{\mu4}$ by operators $P_\mu$, defined by
\begin{equation}
S_{\mu4} = R \; P_\mu                                          \label{4-1}
\end{equation}
with a (dimensionless) number $R$,
inserting (\ref{4-1}) into (\ref{3-11}), dividing both sides by $R^2$ and, 
finally, taking the limit $R\to\infty$. 
Then the commutator (\ref{3-11}) is replaced by
\begin{equation}
[P_\mu, P_\nu] = 0\;.                                          \label{4-2}
\end{equation}
This limit is known as \begin{em}contraction limit\end{em} or 
\begin{em}group contraction\end{em} \cite{iw}.

If $R$ is large but fixed, (\ref{4-2}) will not be completely satisfied, 
but can nevertheless serve as an approximation to (\ref{3-11}). 
In this sense the contraction limit is often used in physical problems.
In this way it will be used in the following. 

The commutation relations (\ref{3-10}), reformulated for  
$S_{\mu\nu}$,
\begin{equation}
[S_{\mu\nu}, S_{\rho\sigma}] =
-i[g_{\mu\rho} S_{\nu\sigma} - g_{\mu\sigma}
S_{\nu\rho} + g_{\nu\sigma} S_{\mu\rho}
- g_{\nu\rho} S_{\mu\sigma}] \; ,                         \label{4-3a}
\end{equation}
define the homogeneous 
\begin{em}Lorentz group\end{em} SO(3,1) as a subgroup of SO(3,2).
They are not changed by the contraction process.
From (\ref{3-12}) the commutation relations of $P_\mu$
with the generators of Lorentz transformations are obtained
\begin{equation}
[S_{\mu\nu}, P_\rho]
= i[g_{\nu\rho} P_\mu - g_{\mu\rho} P_\nu] \;.                   \label{4-3b}
\end{equation}
(\ref{4-2}), (\ref{4-3a}) and (\ref{4-3b}) are the
commutation relations of the \begin{em}Poincar\'e group\end{em} P(3,1).
The quasi-continuous quantum numbers of SO(3,2) are now replaced by the
continuous spectrum of $P_\mu$.

The contraction limit delivers an approximate description of
the symmetry group SO(3,2) by the Poincar\'e group, which is valid for
infinitesimal operations generated by $S_{\mu4}$. Because of the rescaling, 
these correspond to finite transformations generated by $P_\mu$.

Eigenstates $|P\rangle$ of $P_\mu$ have to be considered as approximations 
to the states of clusters of Dirac spinors. 
More accurately, they serve as approximations within an infinitesimal 
small neighbourhood $\cal{N}$ of an arbitrarily chosen origin $\cal{P}$,
with respect to transformations generated by $S_{\mu4}$.

These eigenstates can be used to construct new states $|X\rangle$ that are 
\begin{em}localised\end{em} in space-like directions
\begin{equation}
|X\rangle := (2\pi)^{-3/2} \int\limits^\infty_{-\infty}\! d^3P
\: e^{i x^\mu P_\mu}\, |P\rangle\;.                             \label{4-4}
\end{equation}
When applied to these states, $P_\mu$ generate translations by
4-vectors $a^\mu$
\begin{equation}
e^{ia^\mu P_\mu} \, |X\rangle = |X + a\rangle \;.               \label{4-5}
\end{equation}
The 4-vectors $a^\mu$ span a 4-dimensional real vector space 
with Minkowskian metric, which, with respect to P(3,1), has the transformation
properties of \begin{em}space-time\end{em}.

Obviously, space-time is obtained as a property of macro-states. 
Therefore, in the absence of macro-objects, space-time is not defined.

The quantum mechanical consequences of definition (\ref{4-4}) are
the well-known commutation relations of momentum and position. 

In terms of eigenstates of $P_\mu$, the spin network can now be described in
the following way.
There are states of representations of the Poincar\'e group P(3,1), which can 
be linked to form common states of new representations of P(3,1). 
In this way a network of momentum states is formed, which spreads out over 
the 4-momentum space. 
Momentum space is obtained as a space of \begin{em}possible\end{em}
values of quantum numbers that label eigenvectors of $P_\mu$. 
Through the coverage by a network of momentum states it materialises as a 
physical entity.
A similar consideration applies to the complementary picture defined by
localised states (\ref{4-4}).

Two macro-states can be \begin{em}entangled\end{em} to form a product state 
with momentum $P$
\begin{equation}
|P\rangle = \int d^3q\;c(q)\,|P_1 - q \rangle\;|P_2 + q \rangle\;.\label{4-6}
\end{equation}
The two macro-objects are then linked by the ``exchange"' of three components 
of momentum $q$.
They form a coherent system, because their state cannot be written as a 
product of two states.

Obviously, a macro-object can be entangled also with two or even three other
objects as indicated by
\begin{equation}
|P\rangle = \int dq_1 dq_2 dq_3 \;c(q)\,
    |P_1 - (q_1, q_2, q_3) \rangle   \;
    |P_2 + (q_1,   0, q_3) \rangle   \;  
    |P_3 + (  0, q_2,   0) \rangle                          \label{4-7}
\end{equation}
and
\begin{equation}
|P\rangle = \int dq_1 dq_2 dq_3\;c(q)\,
    |P_1 - (q_1, q_2, q_3) \rangle \;  
    |P_2 + (q_1,   0,   0) \rangle \;    
    |P_3 + (  0, q_2,   0) \rangle \;      
    |P_4 + (  0,   0, q_3) \rangle     \;.                    \label{4-8}
\end{equation}

In all cases the three components of momentum act like chemical valences, 
allowing the first subsystem to link to a maximum of three other subsystems.

Entanglement of states is a typical quantum mechanical phenomenon, with 
impressive applications in quantum computing, encryption and even
``teleportation" \cite{az}. 
Below it will be shown that the type of entanglement, as defined by 
(\ref{4-6}), (\ref{4-7}) or (\ref{4-8}), significantly influences the
formation of \begin{em}particles in space-time\end{em}.

\section{Particles in space-time}

Consider a macro-object of $N$ Dirac spinors with total momentum $P$.
Then add another spinor, described by a 4-component vector $\psi_s$ 
to the system.
Within the Poincar\'e approximation the combined state describes a 
macro-object of $N+1$ spinors with a slightly different momentum $P + p$.
Therefore, \begin{em}relative to the macro-object\end{em} the spinor has a 
momentum $p$.

Now apply a Lorentz transformation to the combined system.
Then both momenta $P$ and $p$ transform as 4-vectors, whereas $\psi_s$
transforms as a Dirac spinor. 
From the Dirac theory of electrons it is known that under such a 
transformation the product $\gamma^\mu p_\mu$ is invariant.

A numerical value can easily be assigned to this invariant.
The energy operator of the combined system is defined as the approximation  
to the operator $L_{04} + \frac{1}{2}\gamma_0$. 
With a spinor state $\psi = (1,0,0,0)$ and $\gamma_0 = diag(1, 1, -1, -1)$,
the value $\frac{1}{2}$ is obtained as the contribution of the spinor to the 
energy of the macro-object, in units of SO(3,2) quantum numbers.
Hence, the state $\psi$ of a Dirac spinor and its contribution $p$ to the 
momentum of a macro-object must obey the Dirac equation
\begin{equation}
(\gamma^\mu p_\mu - m) |\psi_s\rangle = 0  \; ,                 \label{5-5}
\end{equation}
with mass $m = \frac{1}{2}$, if $p$ is measured in the same units as $m$.

The mass $m$ is dimensionless. It is defined in ``units of quantum numbers".
Values in units of a mass will be obtained, when a ``standard" i.e. the mass 
of a special particle is determined, to which all further measurements of 
masses are refered. 
Obviously, a candidate for such a reference has just been found.

Since the value of the momentum $p$ and the associated spinor state $\psi_s$ 
do not depend on any specific property of the macro-object, 
it makes sense to label the spinor state with this value of $p$ and write 
$\psi_s(p)$. 
This is more than a formal act. Since these spinor states, when attached
to macro-states, add the momentum $p$ to the macro-object,
it can be said that, in the environment of macro-objects, these spinor states 
``carry" a momentum of $p$.
Recoupling a spinor state from one macro-object to another, therefore,
describes the transport of 4-momentum.

The new degrees-of-freedom, expressed by the momentum $p$, of course, do 
not belong to the spinor itself. 
Nor do they stand for degrees-of-freedom within the linking of the spinor 
to the macro-object. 
What they express, are degrees-of-freedom in the linkage of the macro-object 
to the observing system. 
This is evident from the way the spectrum of $p$ has been generated by 
applying a Lorentz transformation to the combined system of spinor and
macro-object.
Alternatively, the Lorentz transformation could have been applied to the 
observing system.
Generally spoken: the momentum degrees-of-freedom describe the 
degrees-of-freedom in embedding the spinor into a network of macroscopic 
clusters of spinors.
A descriptive way of looking at $p$ is, obviously, to understand it as an 
\begin{em}orbital momentum\end{em} relative to the observing system.

The way space-time properties of a Dirac spinor have been introduced may
appear rather coarse. 
However, as long as the approximate Poincar\'e invariance is valid, 
this, at least, delivers an adequate description of such processes 
that are determined by the exchange of energy-momentum between a single 
Dirac spinor and macro-objects. 
In other words, a description of ``free" particles has been obtained. 
Of course, this model is not adequate for the description of,
i.e. two ``interacting" particles without the inclusion of 
suitable ``higher order corrections". Steps in this direction have
been described in \cite{ws1,ws2}.

What can be said about the statistics of these particles?
By definition, spinors cannot be distinguished except by the spin variable.
Therefore, when two spinors are interchanged, their common state does 
not change, except for a phase factor.
The interchange of spinors in space-time can be performed by a rotation
of $180$ degrees with respect to a symmetry axis of the system.
A simple calculation delivers a phase factor $i$ for each spinor. 
Then the total state is changed by a factor of $-1$.
Therefore, the particles are subjected to Fermi statistics.
This derivation of Pauli's principle goes back to
A. A. Broyles \cite{aab} in 1976. Later it was used, obviously independently, 
by Feynman and Weinberg \cite{fw}.

Summarising it can be stated that spinors, when described in relation to
macro-objects, show properties of spin-1/2 fermions in space-time.

\section{Mass spectrum}

In the last section, a mass-scale was found, which will now be used 
to determine mass relations with more complex configurations.

Assume that the observing system is linked by forms (\ref{4-7}) or 
(\ref{4-8}) to two or three macro-objects, respectively, with an additional 
spinor linked to this aggregate of macro-objects.
In order to compare the multi-object configurations to the single
object situation treated obove, they must be reduced to, eventually a direct 
sum of, representations of the Poincar\'e group.
Then in each representation, a partial mass will be found with the same value 
of $1/2$.
The effective mass is then determined by evaluating the sum of 
representations within the direct sum.
Although each representation may contribute a different momentum, 
the number of representations should not depend on the momenta involved.
Therefore, the determination of the effective mass, primarily means finding
all representations of P(3,1) that contribute to the multi-object
situation. 
To this aim, first a SO(3,2) representation will be decomposed into a set of
SO(3,1) representations.

Let $S$ denote the group of all SO(3,2) transformations. 
Let $L$ denote the group of Lorentz transformations of SO(3,2), which forms a 
subgroup of $S$, and let $P$ denote the transformations of the (approximate) 
Poincar\'e group P(3,1).

Consider a macro-object with state $|\Phi\rangle$ in a SO(3,2)-symmetric 
Hilbert space $H_S$.
Assume that in the neighbourhood $\cal{N}$ of the origin $\cal{P}$ this 
state is approximated by a momentum eigenstate.
When all Lorentz transformations $L$ are applied to this state,
a Hilbert space $H_L$, ``local" to $\cal{N}$, as a subspace of $H_S$, 
is obtained. 

If a (finite) transformation $s \in S, s \not\in L$ is applied to a state 
of $H_L$, a new state is generated, which is not in $H_L$. 
Therefore, by applying transformations of the \begin{em}coset\end{em} $Ls$, 
a non-equivalent Hilbert space $H^s_L$, local to $s\cal{N}$, is obtained.
There is a one-to-one relation between cosets $Ls$ and Hilbert spaces
$H^s_L$.
The set of all cosets $Ls$ generates the total Hilbert space $H_S$.

The set of cosets forms a \begin{em}homogeneous space\end{em} $S/L$,
where $S$ acts transitive on this space and $L$ is the isotropy group
of the origin $\cal{P}$;
the projection $\pi:\,S \rightarrow S/L$ makes $S$ a principle bundle
on $S/L$ with fiber $L$ \cite{tr}. 

Adding up all the non-equivalent $H^s_L$ means an integration over the
homogeneous space $S/L$.
The integral delivers a decomposition of $H_S$ in terms of $H^s_L$
\begin{equation}
H_S = \int d\Omega\;H^s_L = \int ds\,\frac{d\Omega}{ds}\;H^s_L\;, \label{6-1}
\end{equation}
where $d\Omega$ is the infinitesimal volume element in $S/L$.
The Jacobian $d\Omega/ds$ is a measure of the number of 
non-equivalent Hilbert spaces $H^s_L$ obtained by an infinitesimal 
transformation $ds$.

Integrating a constant $C$ over $S/L$ results in the product of $C$
with the volume $V$ of $S/L$
\begin{equation}
\int C\;d\Omega \; =\; C\int ds\,\frac{d\Omega}{ds}\; =\;C\;V(S/L)\;.\label{6-5}
\end{equation}
With a properly chosen parameterisation, such that the Jacobian does not depend
on $s$ and $\int \! ds = 1$, V(S/L) is identical to the Jacobian in (\ref{6-1}).
The volume of $S/L$ has been calculated in \cite{ggm1} 
\begin{equation}
V(S/L) = \frac{16 \pi}{3}\;.                              \label{6-7}
\end{equation}

Volumes of homogeneous spaces were systematically calculated by 
L. K. Hua \cite{hl,lkh}.
They have been used with some success in semi-empirical mass formula 
for more than three decades \cite{aw1,fds}.

Recently G.~Gonz\'alez-Mart\'in \cite{ggm1,ggm} (G-M in the following)
has obtained mass relations, based on an universal structure group SL(4,R).
G-M's idea is that the structure group describes a ``substrate", from which
particles are generated as ``excitations" with certain symmetric and
topological properties, which are associated with subgroups of the structure
group.

G-M has found a mass formula for the three massive leptons
\begin{equation}
m_n = 4\pi \left(\frac{16 \pi}{3}\right)^n \; m_e 
\hspace{1cm}  n = 1, 2 \;\; ,                                \label{6-8}
\end{equation}
where $m_e$ is the electron mass and $m_1$ stands for 
the myon mass, $m_2$ to the tauon mass.
With the experimental electron mass of $0.5109989$ MeV, G-M obtains
$m_\mu = 107,5916$ MeV and $m_\tau = 1770,3$ MeV.
(The experimental values are $105,658$ and $1776,99$.)

It will be shown that the decomposition (\ref{6-1}) of $H_S$ leads directly
to an identical mass formula.
The proof is as follows.

In the case of two or three macro-objects, the Hilbert space $H_S$ is
obtained from the direct product of individual Hilbert spaces
$H^{(1)}_S, H^{(2)}_S$, and eventually $H^{(3)}_S$. 
With the decomposition (\ref{6-1}) of each Hilbert space,
the integrals contain products of volume factors (Jacobians) $V(S/L)$.
The following factors correspond to one, two and three 
macro-objects, respectively,
\begin{equation}
\left(\frac{16 \pi}{3}\right) \mbox{,} \;\;
\left(\frac{16 \pi}{3}\right)^2  \;\; \mbox{ and } \;\;
\left(\frac{16 \pi}{3}\right)^3.                         \label{6-9}
\end{equation}

Since the effective spinor mass is determined relative to macro-states 
in $\cal{N}$, only the Hilbert space $H_L$, local to $\cal{N}$,
will contribute.
Therefore, when $H_S$ is decomposed into local Hilbert spaces $H^s_L$,
all $H^s_L$ with $s\not\in L$ can be dropped.
This step eliminates one integration over $S/L$ and, therefore, divides 
the Jacobians in (\ref{6-9}) by a factor of $V(S/L)$.

Next remember that spinors are attached to macro-states 
of representations of P(3,1), rather than of SO(3,1).
Representations of P(3,1) are obtained from representations of SO(3,1)
by applying infinitesimal transformations $t \in S$.
By adding transformations $t$ to $L$, again cosets $Lt$ can be defined, 
forming a homogeneous space $P/L$ with a volume \cite{ggm}
\begin{equation}
V(P/L)\;=\;V(U(1))\;=\;4\pi \;.                          \label{6-9a}
\end{equation}
In multi-object cases, this adds an integration over the homogeneous
space $P/L$ with a Jacobian of $V(P/L)$.

A scaling factor, including both steps, is then given by
\begin{equation}
\frac{V(P/L)}{V(S/L)}\;=\;4\pi \frac{3}{16\pi}\;=\;3/4\;. \label{6-10}
\end{equation}

In the one-object case, a modification of this factor is required. 
Here, infinitesimal $t$ applied to $H_L$ are approximated by translations.
These do not affect momentum eigenstates, except by a phase factor.
This means, the addition of a translation $t$ to $L$ generates  
a Hilbert space $H^t_L$ that is identical to $H_L$.
Therefore, in the one-object case, $V(P/L)$ has to be replaced by $1$. 
Instead of (\ref{6-10}), the scaling factor
\begin{equation}
\frac{1}{V(S/L)}\;=\;\frac{3}{16\pi}                     \label{6-12}
\end{equation}
must be used.

In multi-object cases, an infinitesimal $t$, applied to an individual
$H^{(i)s}_L$, generates a mixture of translations, rotations and boost 
operations in $H^{(i)s}_L$. 
This is a consequence of the commutation relations between 
the generators of $t$ and (finite!) transformations $s$.
Such changes of the individual states, in general, cannot be reduced 
to translations in $H^{(i)}_L$.
Therefore, in these cases the factor (\ref{6-10}) does apply.

Multiplying the terms in (\ref{6-9}) by the appropriate factors 
(\ref{6-12}) and (\ref{6-10}), results in the following volume factors
\begin{equation}
1  \mbox{ , } \;\;
4\pi \left(\frac{16 \pi}{3}\right)  \;\; \mbox{ and } \;\;
4\pi \left(\frac{16 \pi}{3}\right)^2  .                 \label{6-14}
\end{equation}
This means, in $\cal{N}$ the entanglement of two and three macro-states 
generate a direct sum of non-equivalent Hilbert spaces $H_P$, with a 
multiplicity determined by an integral over a parameter space with a 
Jacobian given by (\ref{6-14}).

The spinor, linked to a multi-object configuration, is 
linked to every momentum eigenstate in each Hilbert space $H_P$.
It delivers a contribution to the mass of $m_e = 1/2\,$ in each $H_P$.
The total effective spinor mass is then determined by an integral over a 
volume corresponding to one of the volume factors of (\ref{6-14}).
But since the spinor mass is the same in each $H_P$, the integral can
be replaced by a multiplication of $m_e$ by the volume factors of 
(\ref{6-14}). 
This reproduces G-M's mass relations (\ref{6-8}).

The mathematical steps performed before can be illustrated in the
following way:
Starting from a neighbourhood $\cal{N}$ of point $\cal{P}$,
a macro-state is selected and Lorentz transformations in $\cal{N}$ 
are applied to this state.
This generates a Hilbert space $H_L$.
In $\cal{N}$ the states of $H_L$ are locally approximated by momentum 
eigenstates in the tangential plane at $\cal{P}$.
A transformation $s$ generates a new Hilbert space $H^s_L$. 
In $s\cal{N}$ its states can again be approximated by momentum eigenstates
in the tangential plane of $s\cal{P}$.
In this way a coverage of $H_S$ by a set of $H^s_L$ is obtained.
The coverage allows for a piecewise approximation of any state of 
$H_S$ by (linear combinations of) momentum eigenstates in 
neighbourhoods $s\cal{N}$ of $s\cal{P}$.
Any state is thereby decomposed into a direct sum of ``local" states in 
$H^s_L$.
By forming a direct product of two such states, each local state of 
one object is multiplied by each local state of the other.
Then all product states are collected into matching tangential 
planes. 
Finally, a spinor is linked to the resulting momentum eigenstates 
in the tangential plane at $\cal{P}$.
The multiplicity of these momentum eigenstates is given by factors
of $(\ref{6-14})$, which result from the piecewise decomposition
(\ref{6-1}).

The fact that the mass relations agree with experimental data suggests 
an identification of the three configurations with a representation of 
massive leptons.
The actual existance of the lepton mass spectrum then delivers strong
experimental support for the concept of a spin network with a basic 
SO(3,2) symmetry, as presented here.

The preceding statements, in principle, apply also to the massless 
solutions of the Dirac equation. 
Therefore, each massive lepton is accompanied by a massless, neutrino-like 
lepton.
Hence, the three types of linkage between macro-objects, which are provided
by the three components of momentum, generate   
\begin{em}three families of leptons\end{em}.

\section{Conclusion}

In this article a rudimental quantum mechanical theory of space-time has been 
presented.
Minkowskian space-time has been obtained in a rather unspectacular way, as a 
flat-space approximation for quasi-classical clusters of Dirac spinors.
This approximation is valid for the neighbourhood of any given point in 
space-time. 
It is evident from the structure of the basic symmetry group that for larger
translations, both in space and time, the flat-space approximation will no 
longer be valid.
This means, higher approximations will result in a curved space-time 
structure.
But even then, space-time will very likely have to be derived 
as a property of quasi-classical clusters of spinors.
This will be the subject of further investigations.

In a quantum theory of space-time, one may expect to find a quantised 
analogue of the classical space-time structure.
The network of Dirac spinors seems to serve this purpose, though not
in space-time, but in the complementary space of 4-momentum.
However, there is no fine-structure of space-time itself, as might have 
been expected, for example some kind of ``quantum foams" or indications of 
``quantum loops".
The reason is quite simple: space-time has been derived as a 
\begin{em}property of quasi-classical structures\end{em}, and not as an 
independent physical entity.
It is a space made up by the values of parameters, which describe finite 
transformations of a symmetry group,
applied to states with quasi-continuous quantum numbers.
Obviously, it does not make sense to ask for any underlying fine-structure 
of plain parameters.
Of course, the spinor elements, which make up the quasi-classical structures,
form a fine-structure. 
But this certainly does not imply any fine-structure of space-time itself, 
but rather of the matter that is embedded in space-time.
In this sense, individual spinors are manifestations of ``internal quantum 
numbers", whereas large clusters of spinors develop space-time as a collective
property. 

For more than three decades, spin networks have been considered a promising 
basis for a fundamental theory of space-time.
Unfortunately, such a theory has been sought for at scales of the Planck
length.
This has made it virtually impossible to obtain any seminal support from 
experimental observations.
The approach, presented in this article, has introduced a spin network at
the sub-atomic level,
with the clear advantage that its results can directly be compared with
the experiments of particle physics.
In this sense, the actual observation of the lepton mass spectrum, 
delivers experimental evidence in favour of the approach and its 
interpretation presented here.

\renewcommand{\baselinestretch}{1.1}


\begin{thebibliography}{99}

\bibitem{rp} R.~Penrose, ``Angular momentum: an approach to
combinatorial space-time", in: \begin{em}Quantum Theory and Beyond\end{em},
ed. Ted Bastin (Cambridge University Press, Cambridge, 1971).

\bibitem{rpwr} R.~Penrose, W.~Rindler, 
\begin{em}Spinors and space-time\end{em}, p. 43,
(Cambridge University Press, Cambridge, 1988).

\bibitem{jcb} J. C. Baez, ``An introduction to spin foam models of
quantum gravity and BF theory", in: \begin{em}Geometry and Quantum
Physics\end{em}, eds. H. Gausterer and H. Grosse, (Springer, Berlin, 2000);
available as gr-qc/9905087.

\bibitem{iw} E. In\"on\"u and E. P. Wigner,
Proc. Nat. Acad. Sci. USA \bfseries 39\mdseries, 510 (1953).

\bibitem{az} A. Zeilinger et al., Nature \bfseries 390\mdseries, 575 (1997).

\bibitem{ws1} W. Smilga, ``Spin foams, causal links and geometry-induced
interaction", in: \begin{em}Frontiers in General Relativity and Quantum
Cosmology Research\end{em}, ed. V. H. Marselle, 
(Nova Science Publishers, New York, 2006);
available as hep-th/0403137.

\bibitem{ws2} W. Smilga, ``Quantum Electrodynamics based on a Superselection 
Rule", available as hep-th/0508152.

\bibitem{aab} A. A. Broyles, Am. J. Phys. \bfseries 44\mdseries, 340 (1976).

\bibitem{fw} R. Feynman and S. Weinberg, ``The Reason for Antiparticles",
in: \begin{em}Elementary Particles and the Laws of Physics\end{em},
(Cambridge University Press, New York, 1987).

\bibitem{tr} T. Rowland, ``Homogeneous Space", from \begin{em}MathWorld\end{em}
-- A Wolfram Web Resource, created by Eric W. Weisstein.

\bibitem{ggm1} G. Gonz\'alez-Mart\'in, ``p/e Geometric Mass Ratio",
Reporte SB/F/278-99, Univ. Sim\'on Bol\'ivar (1999);
available as physics/0009066.

\bibitem{hl} L. K. Hua and K. H. Look (= Lu, Qi-keng),
Scientia Sinica \bfseries8\mdseries, 1031-1094 (1959).

\bibitem{lkh} L. K. Hua, \begin{em}Harmonic Analysis of Functions of
Several Complex Variables in the Classical Domains\end{em},
(American Mathematical Society, Providence, 1963).

\bibitem{aw1} A. Wyler,
C. R. Acad. Sc. Paris \bfseries271A\mdseries, 180 (1971).

\bibitem{fds} F. D. Smith, Jr.,
Int. J. Theor. Phys. \bfseries24\mdseries, 155 (1985);
\bfseries25\mdseries, 355 (1986).

\bibitem{ggm} G. Gonz\'alez-Mart\'in, ``Lepton and Meson Masses",
Reporte SB/F/304.4-02, Univ. Sim\'on Bol\'ivar (2003);
available as physics/0405094.

\end{thebibliography}
\end{document}